\documentclass[sigconf]{acmart}

\usepackage{amsfonts}
\usepackage{graphicx}
\usepackage{amsmath}
\usepackage{multirow}
\usepackage{balance}
\usepackage[ruled,vlined,linesnumbered]{algorithm2e}
\newtheorem{myDef}{\textbf{Definition}}

\usepackage{textcomp}

\AtBeginDocument{%
  }

\copyrightyear{2025}
\acmYear{2025}
\setcopyright{acmlicensed}\acmConference[KDD '25]{Proceedings of the 31st ACM SIGKDD Conference on Knowledge Discovery and Data Mining V.1}{August 3--7, 2025}{Toronto, ON, Canada}
\acmBooktitle{Proceedings of the 31st ACM SIGKDD Conference on Knowledge Discovery and Data Mining V.1 (KDD '25), August 3--7, 2025, Toronto, ON, Canada}
\acmDOI{10.1145/3690624.3709387}
\acmISBN{979-8-4007-1245-6/25/08}
\begin{document}

\title{Breaker: Removing Shortcut Cues with User Clustering for Single-slot Recommendation System}

\author{Chao Wang}
\affiliation{%
  \institution{Meituan Group}
  \city{Shanghai}
  \country{China}}
\email{wangchao153@meituan.com}

\author{Yue Zheng}
\affiliation{%
  \institution{Meituan Group}
  \city{Shanghai}
  \country{China}}
\email{zhengyue08@meituan.com}

\author{Yujing Zhang}
\affiliation{%
  \institution{Meituan Group}
  \city{Shanghai}
  \country{China}}
\email{zhangyujing06@meituan.com}

\author{Yan Feng}
\affiliation{%
  \institution{Meituan Group}
  \city{Shanghai}
  \country{China}}
\email{fengyan14@meituan.com}

\author{Zhe Wang}
\authornote{Corresponding author.}
\affiliation{%
  \institution{Meituan Group}
  \city{Beijing}
  \country{China}}
\email{wangzhe36@meituan.com}

\author{Xiaowei Shi}
\affiliation{%
  \institution{Meituan Group}
  \city{Beijing}
  \country{China}}
\email{shixiaowei02@meituan.com}

\author{An You}
\affiliation{%
  \institution{Meituan Group}
  \city{Beijing}
  \country{China}}
\email{youan@meituan.com}

\author{Yu Chen}
\affiliation{%
  \institution{Meituan Group}
  \city{Beijing}
  \country{China}}
\email{chenyu17@meituan.com}
\renewcommand{\shortauthors}{Chao Wang et al.}

\begin{abstract}
In a single-slot recommendation system, users are only exposed to one item at a time, and the system cannot collect user feedback on multiple items simultaneously. Therefore, only pointwise modeling solutions can be adopted, focusing solely on modeling the likelihood of clicks or conversions for items by users to learn user-item preferences, without the ability to capture the ranking information among different items directly. However, since user-side information is often much more abundant than item-side information, the model can quickly learn the differences in user intrinsic tendencies, which are independent of the items they are exposed to. This can cause these intrinsic tendencies to become a shortcut bias for the model, leading to insufficient mining of the most concerned user-item preferences. To solve this challenge, we introduce the Breaker model. Breaker integrates an auxiliary task of user representation clustering with a multi-tower structure for cluster-specific preference modeling. By clustering user representations, we ensure that users within each cluster exhibit similar characteristics, which increases the complexity of the pointwise recommendation task on the user side. This forces the multi-tower structure with cluster-driven parameter learning to better model user-item preferences, ultimately eliminating shortcut biases related to user intrinsic tendencies. In terms of training, we propose a delayed parameter update mechanism to enhance training stability and convergence, enabling end-to-end joint training of the auxiliary clustering and classification tasks. Both offline and online experiments demonstrate that our method surpasses the baselines. It has already been deployed and is actively serving tens of millions of users daily on Meituan, one of the most popular e-commerce platforms for services.
\end{abstract}



\maketitle

\section{Introduction}
\begin{figure}[htbp]
  \centering
  \includegraphics[width=\linewidth]{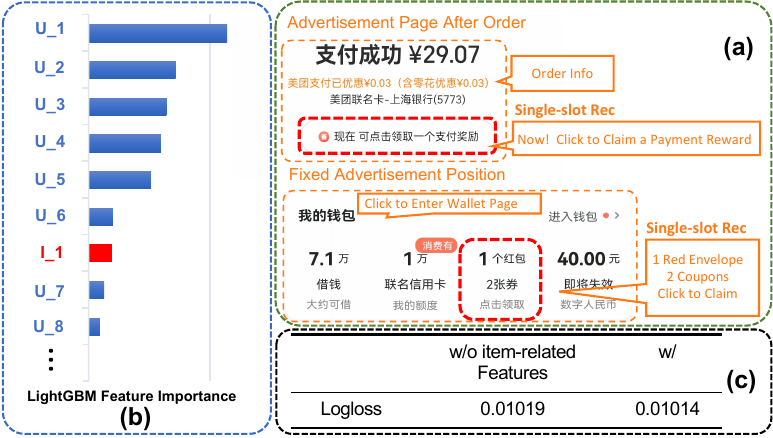}
  \caption{(a) An example of single-slot recommendation. The recommendation system presents various copywriting expressions within the red dotted line to attract users to click. (b) Our scenario shows the feature importance scores in the trained LightGBM model. Among them, red represents item-related features and blue represents user-related features. (c) There are slight differences in model performance with the addition and removal of item-related features.}
  \label{fig:intro}
  \Description{}
\end{figure}

In recent years, recommendation systems have seen explosive growth and widespread adoption across diverse fields, including e-Commerce \cite{zhou2018deep, chen2022extr, lyu2022see}, social media \cite{el2022twhin, wang2021masknet, huang2019fibinet, he2017neural}, digital streaming \cite{covington2016deep, zhao2023disentangled, huang2021sliding}, financial services \cite{cui2019hidden, xi2021modeling, li2020spending, yu2021joint}, and more.
The essence of a recommendation system lies in predicting user preferences based on user-item impression data, with the ultimate goal of achieving business objectives such as increasing user engagement through user clicks\cite{lyu2022see, wang2021masknet} or prompting users to complete multi-step conversions \cite{xi2021modeling}.
The form of recommendation systems can be divided into two categories: single-slot recommendation and list-based recommendation.
Single-slot recommendation, as shown by the red dotted box in Figure \ref{fig:intro}(a), refers to a type of recommendation system that recommends only one item at a time to the user, so \textit{the system cannot collect user feedback on different items simultaneously}, leading to \textbf{only pointwise scoring functions can be used in modeling user preferences}.
Consequently, the modeling of user preferences is limited to the indirect inference of the likelihood of clicks or conversions between the user and the recommended items without the ability to capture the ranking information among different items directly.

Existing research \cite{xie2023dewll,lyu2022see} offers an alternative interpretation of user clicks, categorizing them into \textbf{intrinsic tendencies} and \textbf{user-item preferences}. The former reflects users' natural proclivity or passivity, corresponding to their propensity to click or not click on any given item. The latter pertains to users' preferences for specific items, leading them to click only on certain items, which is what the recommendation system is most concerned about.
\cite{geirhos2020shortcut, ScimecaOCPY22} points out that deep neural networks (DNNs) tend to exhibit a shortcut bias, where they learn the easiest possible solution to a problem.
In a single-slot recommendation system,  \textit{the user side often has a much larger number of features than the item side, providing richer information}, which may lead to \textbf{the intrinsic tendencies of the user to be shortcut cues in the DNN with a pointwise scoring function}.
Furthermore, single-slot recommendation systems are commonly used in marketing scenarios where items are driven by business objectives, such as enticing users to try specific services or products. These items are manually configured by the operations and business analysis teams, resulting in a limited selection of items available for recommendation \cite{lyu2022see,xi2021modeling,wang2023multi}. This restricted item pool can limit the exploration of user interests and make it challenging for the model to learn user preferences.

To further elaborate on this issue, we assessed it from \textbf{two aspects: feature importance and the impact of item feature ablation on model performance}. Firstly, we observed the feature importance of the LightGBM model trained in our business.
Figure \ref{fig:intro}(b) shows that the most significant item-related feature ranks only seventh in importance, while user-related features dominate. The importance score of the most significant user feature is six times that of the most significant item feature, indicating a relatively small contribution of item-related features, which might affect the learning of user-item preferences.
We also compared model performance with and without item-related features. Figure \ref{fig:intro}(c) shows that adding item-related features only slightly reduced log loss by $5\times10^{-5}$. This suggests that the model learns user intrinsic tendencies easily but may struggle to learn user-item preferences, making the user's intrinsic tendency a shortcut cue.

Some works have recognized concerns related to user tendencies and preferences, but have not explicitly identified shortcut bias. In \cite{lyu2022see}, the MultiCR model uses multiple classification decoders to model user-item preferences. However, it lacks an inductive bias to remove shortcut cues related to intrinsic user tendencies. We also compared our method with MultiCR in our experiments. \cite{xie2023dewll} filters out clicks with very short dwell times, akin to our scenario where disinterested users can create shortcut biases, limiting the model's ability to uncover user-item preferences.
To solve the challenge, \textbf{we propose a method called Breaker that removes shortcut cues by clustering users}. We introduce an auxiliary unsupervised clustering task for users based on deep embedding clustering.
\textit{In the same cluster, users tend to have similar attributes and thus present similar intrinsic tendencies, making it difficult for the model to distinguish between them}, thus \textbf{forcing the model to learn our most concerned user-item preferences}.
Both online and offline experiments demonstrate the effectiveness.
In summary, our contributions include the following:
\begin{itemize}
    \item As a first attempt, we introduce the shortcut cues existing in single-slot recommendation systems, which makes the model stuck in learning the intrinsic tendency of users, making it difficult to mine the most concerned user-item preferences fully.
    \item We propose a method for removing shortcut cues about the user's intrinsic tendency, called Breaker. To achieve this, we integrate an auxiliary task of user representation clustering with a multi-tower structure for cluster-specific preference modeling. Furthermore, we propose a delayed parameter update mechanism to enhance training efficiency and stability.
    \item Our proposed method has been validated through online experiments, demonstrating its effectiveness. It has been deployed online and is currently serving tens of millions of users on a daily basis.
\end{itemize}

\begin{figure*}[t]
  \centering
  \includegraphics[width=0.92\textwidth]{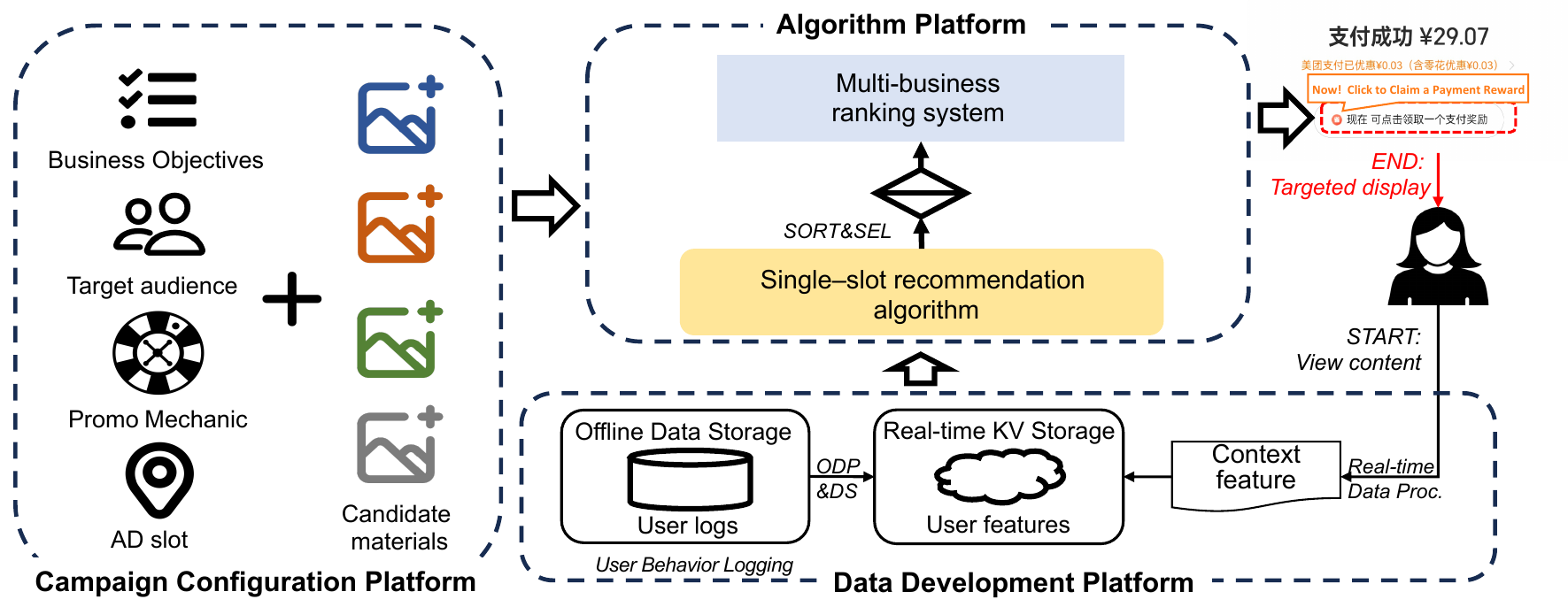}
  \caption{The targeted marketing display recommendation system in Meituan Payment.}
  \label{fig:framework}
  \Description{}
\end{figure*}

\section{The Targeted Marketing Display Recommendation System}
We will provide an overview of the targeted marketing display recommendation system used in our business, where Breaker is implemented. As illustrated in Figure \ref{fig:framework}, the system consists of three main components: the Campaign Configuration Platform, the Data Development Platform, and the Algorithm Platform.

The Campaign Configuration Platform is used by operations personnel to set up marketing campaigns with five elements: business objectives (e.g., user acquisition), target audience (e.g., new or existing users), campaign mechanics (e.g., rewards for specific tasks), marketing positions (placements for campaign deployment), and various display materials. These materials serve as candidate items for the algorithm.

The Data Development Platform includes a distributed storage system for user behavior data, processed into statistical features, and imported into a real-time Key-Value (KV) storage system for on-the-fly model invocation. When a user browses a page, contextual information such as "order amount" is calculated into the features. These features, along with candidate item information, are fed into the Algorithm Platform.

The single-slot recommendation algorithm calculates a preference score for each item and ranks them. Only the top-ranked item will enter the display candidate pool. Since a single slot may be contested by multiple businesses, items are assessed through a multi-business classification system to determine the final display.

\section{Methodology}
We propose Breaker to remove shortcut cues related to users' intrinsic tendencies in user-item preference modeling. Breaker clusters users with similar intrinsic tendencies, increasing the training difficulty on the user side. This forces the model to focus on user-item preferences within each cluster. The mathematical formulation of Breaker is presented below:
\begin{align}
P(Y|U,I) \mathrel{\mathop:}= \sum_{C \in \mathcal{C}} Q(Y|U,I,C)P(C|U),
\end{align}
where $U$ represents the user, $I$ denotes the item, and $P(Y|U,I)$ denotes the likelihood that user $U$ prefers item $I$. $C\in\{\mathcal{C}|C_1,\cdots,C_K\}$ is the user cluster. $Q$ represents the modeling of user preferences for items between users in the cluster $C$. $P(C|U)$ represents the probability of $U$ belonging to $C$.

The structure is illustrated in Figure \ref{fig:model_structure}. Breaker consists of an embedding layer and three modules, i.e., Representation Extraction Module (REM), User Representation Clustering Module (URCM), and Cluster-specific Preference Modeling Module (CPMM).

\subsection{Representation Extraction Module (REM)}
Each training sample is a user-item pair, where the user field comprises features such as age, number of visits in the past $n$ days, browsing behavior, etc., and the item field includes the item ID and potential attribute information.
These features are fed into the embedding layer and each feature is assigned an embedding from the embedding table. Subsequently, we concatenate the feature embeddings to obtain the user embeddings $X_u\in \mathbb{R}^{d_u}$ and the item embeddings $X_i\in \mathbb{R}^{d_i}$.
REM takes user and item embeddings as input and generates user and item representations through a Multi-Layer Perceptron (MLP) structure.
\begin{align}
    E_u&=MLP_u(X_u)\in \mathbb{R}^{d_u^e},\\
    E_i&=MLP_i(X_i)\in \mathbb{R}^{d_i^e},
\end{align}
where $d_u^e$ and $d_i^e$ represent the dimensions of user representation $E_u$ and item representation $E_i$, respectively.

\subsection{User Representation Clustering Module (URCM)}\label{sec:urcm}
After obtaining the user representations, we consider the problem of partitioning $N$ embedded points $\{e_u^i\in E_u\}_{i=1}^N$ into $K$ clusters in the representation space.

\begin{figure}[t]
  \centering
  \includegraphics[width=\linewidth]{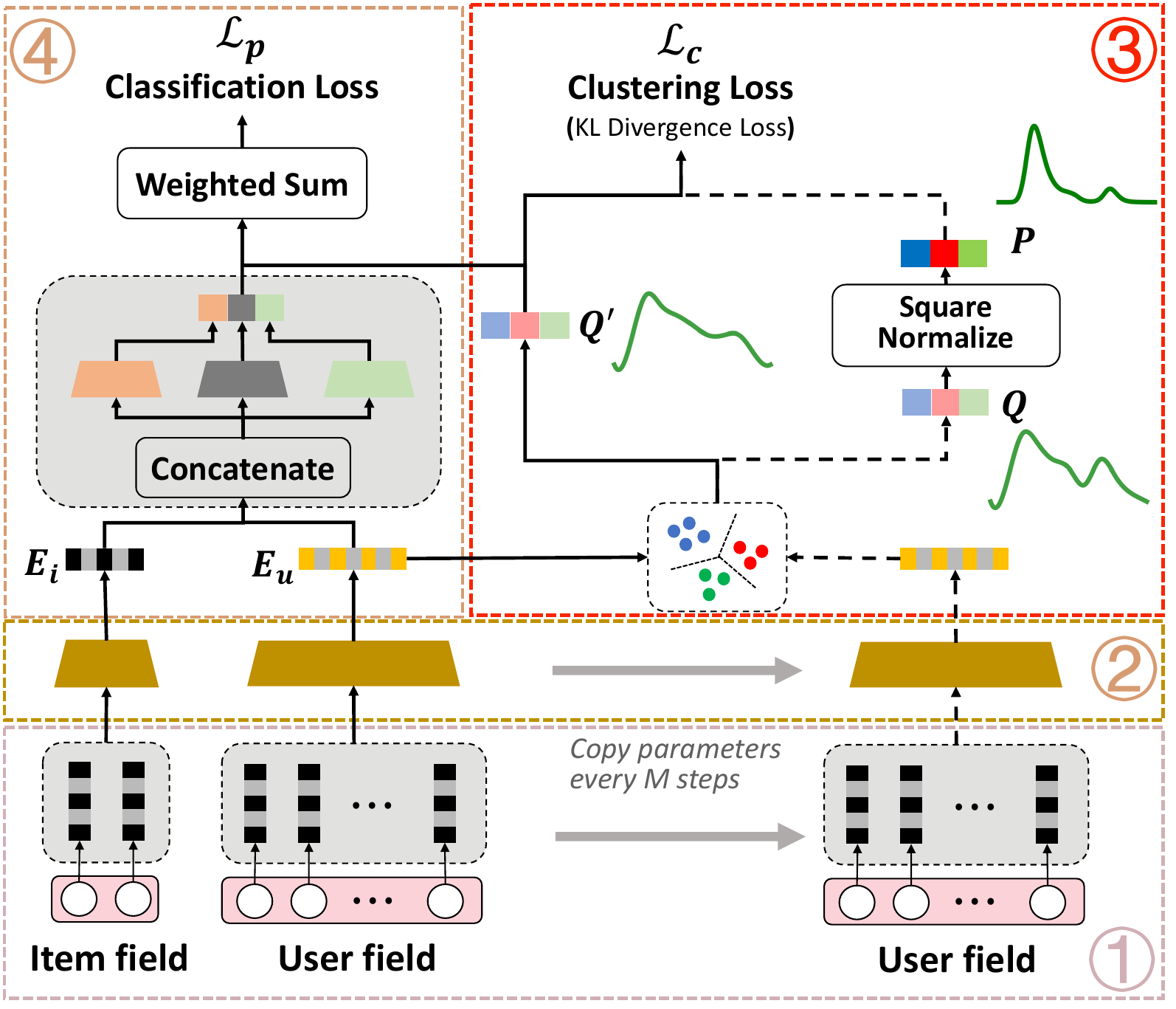}
  \caption{Structure of \textbf{Breaker}, with an \textcircled{\small 1} Embedding Layer, \textcircled{\small 2} Representation Extraction Module, \textcircled{\small 3} User Representation Clustering Module, and \textcircled{\small 4} Cluster-specific Preference Modeling. $\rightarrow$ indicates paths where gradients are generated during training, while $\dashrightarrow$ indicates paths without gradient flow, which are used to generate unsupervised pseudo-labels for the clustering task.}
  \label{fig:model_structure}
  \Description{}
\end{figure}

We draw inspiration from deep embedding clustering \cite{xie2016unsupervised} to construct user clustering modules that enable grouping users with similar characteristics.
As the clustering and classification tasks for estimating user-item preferences have different natures, we employ a delayed parameter update mechanism during the training process to achieve joint training more efficiently.
As shown in Figure \ref{fig:model_structure}, we refer to the classification task as the main network and the clustering task as the target network. The target network outputs pseudo labels to guide the training of the clustering task. These two networks asynchronously share the parameters of the user field embedding and REM.
\begin{myDef}[Cluster Assignment Probability]
Let $Q' = [q'_{ij}] \in \mathbb{R}^{N \times K}$ be a matrix, where $N$ is the number of embedded points $\{e_u^i\in E_u\}_{i=1}^N$ and $K$ is the number of clusters. Each element $q'_{ij}$ of $Q'$ represents the probability of assigning sample $i$ to cluster $j$. Thus, we have $0 \leq q'_{ij} \leq 1$ for all $i \in {1, \ldots, N}$ and $j \in {1, \ldots, K}$, and $\sum_{j} q'_{ij} = 1$ for all $i \in {1, \ldots, N}$.
\end{myDef}

In clustering, Student's $t$ distribution is used as a kernel to measure the similarity between the embedded point $e_u^i$ and the cluster centroid $\mu_j$:
\begin{align}\label{q_ij}
q'_{ij}=\frac{(1+||e_u^i-\mu_j||^2/\alpha)^{-\frac{\alpha+1}{2}}}{\sum_{j'} (1+||e_u^i-\mu_{j'}||^2/\alpha)^{-\frac{\alpha+1}{2}}},
\end{align}
where $\alpha$ is the degree of freedom of the Student’s $t$-distribution. Following \cite{xie2016unsupervised}, we set $\alpha=1$ for all experiments.
\begin{myDef}[Clustering Pseudo-labels]
    Let $P = [p_{ij}] \in \mathbb{R}^{N \times K}$ be an auxiliary target distribution matrix of the same size as $Q'$, which functions to help identify high-confidence assignments and thereby facilitates iterative refinement of the clusters.
\end{myDef}

The similarity calculation $q_{ij}$ is identical to that of $q'_{ij}$ in the main network. The clustering pseudo-labels $p_{ij}$ in the target network are obtained by square normalizing $q_{ij}$.
\begin{align}\label{p_ij}
p_{ij}=\frac{q_{ij}^2/f_j}{\sum_{j'}q_{ij'}^2/f_{j'}},f_j=\sum_i q_{ij}.
\end{align}

The loss of the clustering task is the KL divergence loss between $p_{ij}$ and $q_{ij}'$.
\begin{align}\label{eq:cluster_loss}
\mathcal{L}_c=KL(P||Q')=\sum_{i}\sum_j p_{ij}log\frac{p_{ij}}{q_{ij}'}.
\end{align}

\subsection{Cluster-specific Preference Modeling Module (CPMM)}
After soft clustering, the users within each cluster exhibit similar characteristics and potential intrinsic tendencies.
Compared to learning user-item preferences in the entire sample space, Breaker learns within each user representation space with similar attributes. The first step is to concatenate the user representation $E_u$ and the item representation $E_i$ to form the input of the CPMM.
\begin{align}
E_{in} = [E_u\oplus E_i]\in \mathbb{R}^{d_u^e+d_i^e}.
\end{align}

\begin{figure}[t]
  \centering
  \includegraphics[width=\linewidth]{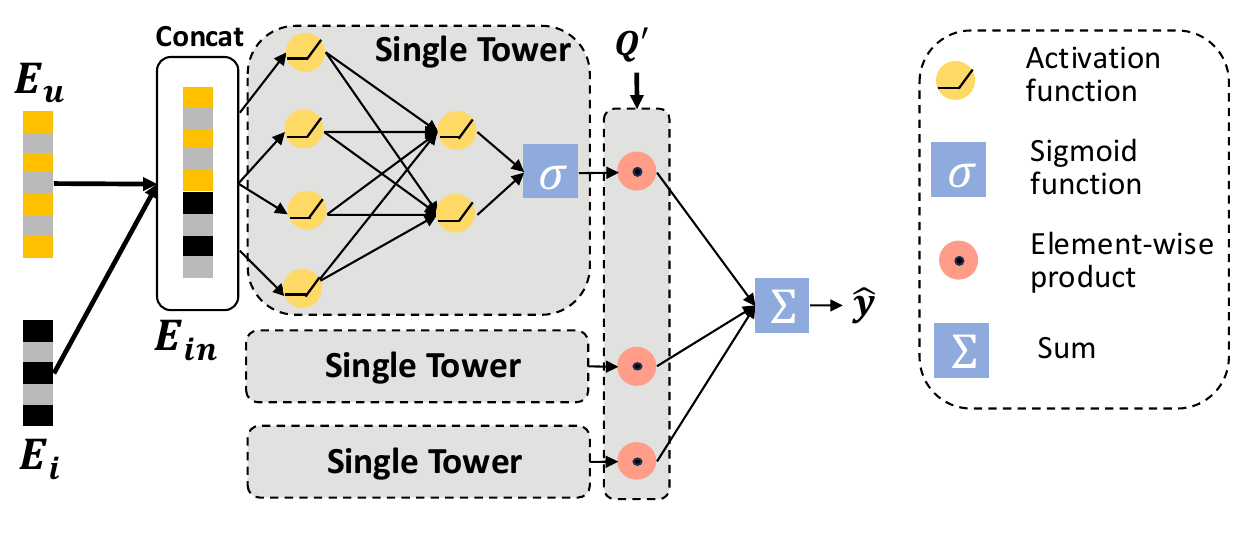}
  \caption{Structure of the Cluster-Specific Preference Modeling Module, which adopts a multi-tower network structure. The output from each tower is multiplied by the user clustering probabilities produced by the User Representation Clustering Module, then aggregated to obtain the user-item preference prediction.}
  \label{fig:cpmm}
  \Description{}
\end{figure}

As shown in Figure \ref{fig:cpmm}, Breaker adopts a multi-tower network structure, where each cluster corresponds to a pointwise classification tower $MLP_k$. The tower output layer uses a sigmoid activation function.
\begin{align}
\hat{y}_c^k = MLP_{k}(E_{in})\in \mathbb{R}^{1}.
\end{align}

To estimate user-item preference, Breaker sums the weighted outputs from each tower, along with the probability of the user's association with each cluster.
\begin{align}\label{eq:final_estimate}
\hat{y}=\sum_{k=1}^K \hat{y}_c^k \times q'_{k}.
\end{align}

Breaker employs the standard negative log-likelihood function as the loss function for the classification task. The final loss of Breaker is the sum of classification loss and clustering loss.
\begin{align}
\label{eq:class_loss}&\mathcal{L}_p=-\frac{1}{N}\sum_{i=1}^N \left(y_i log\hat{y}_i+(1-y_i)log(1-\hat{y}_i)\right),\\
\label{eq:toal_loss}&\mathcal{L}=\mathcal{L}_p+\lambda \mathcal{L}_c,
\end{align}
where the label $y_i$ is a binary value $(0/1)$ that indicates whether user $u$ prefers item $i$, i.e., whether they are likely to click on it or complete multiple rounds of conversion behavior (such as opening a specific service in the marketing customer acquisition scenario), $N$ is the number of samples in the training set. $\lambda$ denotes the loss weight, which is set to 0.1 in our experiments.

\subsection{Parameter Delayed Update Mechanism}
As mentioned in Section \ref{sec:urcm}, Breaker comprises the main network responsible for the classification task and the target network used to cluster users. Due to the significant differences in the nature of the two tasks, training instability may be a problem in end-to-end training.

Drawing inspiration from the achievements of deep reinforcement learning algorithms \cite{mnih2013playing}, we introduce a delayed parameter update mechanism to enhance stability and convergence. This is accomplished by periodically updating the parameters of the target network every $M$ steps while keeping them fixed for a certain duration.
By this method, we alleviate the instability in the training process, which is exacerbated by significant gradient interference stemming from frequent shifts in the task gradient direction during backpropagation.
In the experiment, we empirically set $M$ to be about 10\% of the total steps in an epoch. The general training procedure of Breaker is shown in Algorithm \ref{algo:breaker}.

\subsection{Gradient Analysis}
\subsubsection{Optimization of Clustering Tasks}
We jointly optimize the cluster centroid and DNN parameters. The target network facilitates clustering and influences the user representations $e_u^i$ via $Q'$ and the cluster centroids $\mu_j$.
The gradients of the clustering loss with respect to $e_u^i$ and $\mu_j$ are as follows:
\begin{equation}
    \begin{aligned}
    \frac{\partial \mathcal{L}_c}{\partial e_u^i} &= \frac{\alpha+1}{\alpha}\sum_j \frac{p_{ij}-q_{ij}'}{1+||e_u^i-\mu_j||^2/\alpha} \times( \boxed{e_u^i} -\mu_j) \\
    \frac{\partial \mathcal{L}_c}{\partial \mu_j} &= -\frac{\alpha+1}{\alpha}\sum_i \frac{p_{ij}-q_{ij}'}{1+||e_u^i-\mu_j||^2/\alpha} \times(e_u^i-\boxed{\mu_j}),
    \end{aligned}
\end{equation}
where $\partial \mathcal{L}_c/\partial e_u^i$ facilitated the learning of a representation specifically for clustering. Meanwhile, $\partial \mathcal{L}_c/\partial \mu_j$ have led to improved cluster assignments. We provide a detailed discussion in Section \ref{4_3_1}. Furthermore, the classification loss $\mathcal{L}_p$ also produces gradients with respect to $e_u^i$, encouraging it to be more discriminative.

\subsubsection{Cluster-Driven Parameter Learning in Multi-Tower Preference Modeling}
Multi-tower structure aims to learn user-item preferences within similar user groups, thereby eliminating shortcuts on the user side and enhancing preference learning. Although each tower receives the same intact user representation as input, the integration with the clustering task (as described in Eq.(\ref{eq:final_estimate})) results in different learning directions for the parameters of each tower during optimization. Let $W_k$ denote the parameters of the $k$-th tower. The effect of the clustering task on $W_k$ is as follows:
\begin{align}
 \frac{\partial \mathcal{L}_p}{\partial W_k}=\frac{\partial \mathcal{L}_p}{\partial \hat{Y}}\frac{\partial \hat{Y}^k_c}{\partial W_k}\boxed{Q_k'}.
\end{align}

It is observed that \textbf{the learning of the parameters $W_k$ is predominantly influenced by samples with higher values of $Q_k'$, i.e., users in cluster $k$}. In the extreme case where $Q'_k$ is binary (0/1), each tower learns exclusively from its corresponding user cluster. Since users within the same cluster share similar attributes, this limits the tower's ability to learn diverse user-side information and pushes it toward learning user-item preferences.
\begin{algorithm}[t]
\DontPrintSemicolon
\SetKwComment{Comment}{$\triangleright$\ }{}
 \SetKw{KwBy}{by}
   \SetNoFillComment
    \caption{The training algorithm for Breaker.}\label{algo:breaker}
        \KwIn{training steps $T$, number of clusters $K$, clustering loss weight $\lambda$,  target network update frequency $M$}
        \KwOut{trained neural network}
        \Comment*[l]{$\theta$: parameters of embedding layer, REM and CPMM}
        Initialize the main classification network with $\theta$; \;
        \Comment*[l]{$\theta^{-}$: parameters of user embedding layer and user representation extraction in REM}
        Initialize the clustering target network with weights $\theta^{-}=\theta$;\;
        Calculate user representation $E_u$ through REM with $\theta$;\;
        Initialize $K$ centroids $\mu_k$ for $E_u$ using K-means clustering;\;
        \For{$t=1$ \KwTo $T$ \KwBy $1$}{
            Calculate the probabilities $q_{ij}$ and $q'_{ij}$ within the batch using Eq.(\ref{q_ij}), based on $\theta^{-}$ and $\theta$ respectively;\;
            Calculate clustering pseudo-labels $p_{ij}$ in the target network using Eq.(\ref{p_ij});\;
            Calculate clustering loss $\mathcal{L}_c$ using Eq.(\ref{eq:cluster_loss});\;
            Predict user-item preference $\hat{y}$ using Eq.(\ref{eq:final_estimate}) with tower outputs $\hat{y}_c^k$ and probability $q'_{ij}$;\;
            Calculate the pointwise classification loss $\mathcal{L}_p$ and total loss $\mathcal{L}$ using Eq.(\ref{eq:class_loss},\ref{eq:toal_loss});\;
            Perform a gradient descent step on $\mathcal{L}$ with respect to $\theta$ and cluster centroids $\mu_k$;\;
            \Comment*[l]{Parameter delayed update mechanism for the target clustering network}
            Every $M$ steps, reset $\theta^{-}=\theta$; \;
        }
\end{algorithm}

\section{Experiments}
To verify the effectiveness of Breaker, we have conducted offline experiments and online A/B tests. We first introduce the experimental setup. Finally, we present the experimental results and analysis.

\subsection{Experimental Setup}

\subsubsection{Datasets}
The two datasets for offline experiments (Scenario A and Scenario B) were collected from different ad slot placements in the targeted marketing display services of Meituan Payment. We conducted a randomized controlled experiment (RCT) to collect data, where users were randomly assigned to different buckets based on the result of $hash(userID)$ modulo $n$ (where $n$ is the number of candidate items).
Each bucket of users is exposed to the same item, ensuring a uniform distribution of users for different items and eliminating exposure bias for enhanced model performance.
To minimize business impact, only a small fraction of online traffic was used for data collection, and data was continuously collected over three weeks. The dataset statistics are presented in Table \ref{tab:data_statis}. Scenario A features prominent ad slots that attract highly engaged users, leading to more positive instances, while Scenario B appears in less noticeable slots, resulting in fewer positive instances. Due to the low proportion of positive samples in our scenarios, we set the test set ratio at 15\% for Scenario A and 20\% for Scenario B to ensure a comprehensive evaluation of the method's performance.

\begin{table}[t]
\caption{Statistics of the datasets.}
\label{tab:data_statis}
\begin{tabular}{cccccc}
\toprule
Dataset & \#Users & \#Items & \#Records & \#Positive & \#Feats \\ \midrule
Scenario A & 1,709,938 & 20 & 13,597,208 & 27,523 & 523 \\
Scenario B & 5,416,051 & 18 & 15,125,354 & 5,082 & 501 \\ \bottomrule
\end{tabular}
\end{table}

\begin{table*}[t]
    \caption{The comparison of overall effectiveness and efficiency of various methods across two scenarios. Baseline methods are categorized by their foundational algorithm type, e.g., "FM" represents "FM-based".  "Inf. time" represents the inference time on the test set with identical computational resources. Arrows indicate the direction of improvement where $\uparrow$ means that the higher is better and $\downarrow$ means that the lower is better. The best results are in \textbf{bold}, second best \underline{underlined}.}
    \label{tab:offline_res}
    \begin{tabular}{c|c|cccc|cccc}
    \toprule
    \toprule
    \multirow{2}{*}{ \begin{tabular}[c]{@{}c@{}}Method Class\\ (-based)\end{tabular} } & \multirow{2}{*}{Method} & \multicolumn{4}{c|}{Scenario A}        & \multicolumn{4}{c}{Scenario B}          \\
                           &                        & \#Params (M) & Inf. time (s)$\downarrow$ & Recall@1$\uparrow$ & AER$\uparrow$           & \#Params & Inf. time & Recall@1 &AER    \\ \midrule
    GBDT & LightGBM      & - & \textbf{4.07}         & 0.05040          & 0.002055  & - & \textbf{9.74}        &   0.08259        & 0.0004800           \\ \midrule
    \multirow{1}{*}{FM} & DeepFM        & 1.41 & 19.73         & 0.05536         & 0.002303  & 1.36 & 33.24       & 0.08132        & 0.0004764          \\  \midrule
    \multirow{3}{*}{Cross} & xDeepFM & 40.70 & 6978.30 & 0.05908 & 0.002438 & 37.59 & 9332.62 & 0.08132 & 0.0004975 \\
     & DCN-V2 & 4.13 & 263.49 & 0.05412 & 0.002125 & 3.97 & 351.42 & \underline{0.08895} & 0.0005103 \\
     & EDCN           & 54.80 & 190.09        & 0.05412 & 0.002125   & 50.31 & 281.31       & 0.08641 & \underline{0.0005155}         \\ \midrule
    \multirow{2}{*}{Attention} & AutoInt & 1.42 & 3833.83 & 0.05461 & 0.002194 & 1.37 & 4921.64 & 0.08894 & 0.0004925 \\
     & FRNet      & 4.10 & 1450.91            & 0.05983          & \underline{0.002533}   & 3.94 & 2073.58     & 0.08640          & 0.0004940          \\ \midrule
    \multirow{1}{*}{MLP} & MultiCR        & 0.61 & 20.66           & 0.05735         & 0.002447 & 0.58 & 30.04         & 0.08132        & 0.0004757          \\  \midrule
     \multirow{3}{*}{Ours} & Breaker$_{1-}$      & 1.41 & 12.13    & 0.05288       & 0.002242          & 1.37 & \underline{20.31} & 0.08005         & 0.0004639          \\
     & Breaker$_{2-}$      & 1.41 & 11.96  & \underline{0.06306}          & 0.002505    & 1.37 & 20.41      & 0.08513          & 0.0004978          \\
     & Breaker    & 1.41 & \underline{11.74}            & \textbf{0.06405}           & \textbf{0.002676} & 1.37 & 20.40 & \textbf{0.09530}          & \textbf{0.0005210} \\  \bottomrule \bottomrule
    \end{tabular}
    \end{table*}

\subsubsection{Compared Methods} We compare Breaker with the following representative pointwise methods and its own ablation models, all of which are applicable to single-slot recommendation systems.
\begin{itemize}
    \item \textbf{LightGBM} \cite{ke2017lightgbm}: LightGBM is a gradient-boosting algorithm with excellent performance in modeling tabular data.
    \item \textbf{DeepFM} \cite{guo2017deepfm}: It combines the Factorization Machine and deep learning for feature interactions.
    \item \textbf{xDeepFM} \cite{lian2018xdeepfm}: It combines explicit high-order feature interactions through a Compressed Interaction Network (CIN) with implicit low- and high-order interactions.
    \item \textbf{DCN-V2} \cite{wang2021dcn}: It captures feature interactions by combining a cross network for explicit high-order feature crossing with a deep network for implicit complex feature learning.
    \item \textbf{EDCN} \cite{chen2021enhancing}: It enhances explicit and implicit feature interactions through information sharing.
    \item \textbf{AutoInt} \cite{song2019autoint}: It uses self-attentive neural networks to learn high-order feature interactions automatically.
    \item \textbf{FRNet} \cite{wang2022enhancing}: It can learn context-aware feature representations and can be applied to most prediction models.
    \item \textbf{MultiCR} \cite{lyu2022see}: It proposes a multi-task learning-based model to model the intrinsic tendency to click separately and the preference for different items.
    \item \textbf{Breaker$_{1-}$}: This ablation model differs from Breaker by omitting the clustering auxiliary task.
    \item \textbf{Breaker$_{2-}$}: This ablation model differs from Breaker by excluding the delayed parameter update mechanism.
\end{itemize}

\subsubsection{Evaluation Protocol}
\begin{itemize}
    \item \textbf{Recall@k}: The Recall@k ranking metric is widely used in the realms of information retrieval and recommendation systems. Given the nature of our single-slot recommendation system, our focus is primarily on the performance of the first recommended item, hence set $k=1$.
    \item \textbf{The Average Expected Response (AER)}: To achieve a more consistent evaluation of the methods online, we introduce the AER \cite{zhao2017uplift} in the causal effect modeling. This approach uses the law of total expectation to estimate the expected business objective (e.g., CTR, CVR, or CTCVR) across the entire offline sample space, in contrast to the recall metric which neglects users lacking positive feedback. Notably, this metric provides an unbiased estimation when samples are collected via a randomized controlled trial, consistent with our sample collection methodology.
    \begin{align}
        AER=\frac{1}{|\mathcal{S}|}\sum_{s\in \mathcal{S}}\sum_{i\in \mathcal{I}} y_s \mathrm{1}\{\pi(s_u)=i\}\mathrm{1}\{s_i=i\},
    \end{align}
    where $\mathrm{1}\{\cdot\}$ is the 0/1 indicator function, $s$ denotes a user-item pair $(s_u, s_i)$ sample, $y_s$ is the sample label, $i$ is one of the candidate items $\mathcal{I}$, $\pi$ is the recommendation strategy, where we recommend the item with the highest prediction value through Eq.(\ref{eq:final_estimate}).
\end{itemize}

\subsubsection{Implementation details}
To fairly compare the generalization capabilities of different models, we employ the same hyperparameters across two scenarios, which are determined based on empirical considerations and computational efficiency.
We employ the same neural structure (i.e., 4 layers, 256-64-32-10) for the models that involve MLP to ensure a fair comparison, while the feature extraction layer of Breaker is set to (256-64), with each tower having a layer structure of (32-10). Due to the unique model structure constraints of MultiCR, the user representation layer is set to (64-64), and each item tower is also set to (64-64), to facilitate the computation of dissimilarity loss. We set the embedding dimension to 10 in all experiments. In the Breaker model, we tune $K$ in \{2, 3, 4, 5\}, and eventually set it to 4 in both scenarios. In all experiments, we train the model with SGD using Adam optimizer, with a learning rate of 1e-3, and a batch size set to 2048. We implemented all methods with TensorFlow, and have released the source code of Breaker.\footnote{The code is available at https://github.com/yzheng51/breaker.}

\subsection{Overall Comparison}
\subsubsection{Effectiveness Comparison}
Table \ref{tab:offline_res} summarizes the recall and AER metrics of all models on the offline test set. From these results, we have the following findings.
\begin{itemize}
    \item \textbf{Performance of Baseline Models:} In Scenarios A and B, we observe that different baseline models achieve the best performance. Specifically, FRNet outperforms others in Scenario A, while DCN-V2 leads in Scenario B when focusing on the recall@1 metric. As for AER, EDCN demonstrates superior results. Nonetheless, the results suggest that the performance trends for both recall rate and AER are generally consistent across the scenarios.

    \item \textbf{Breaker Model Performance:} In both scenarios, Breaker surpassed all other baseline models across both metrics. In scenario A, it improves recall@1 by 7.05\% and AER by 5.65\% over the top-performing FRNet. In scenario B, Breaker demonstrated a 7.14\% uplift in recall@1 metric over DCN-V2, along with a 1.07\% increase in AER relative to EDCN.

    \item \textbf{Insights from Ablation Study:} In the ablation study, we compare Breaker with Breaker$_{1-}$, a multi-tower structure with equivalent parameters. The results show that Breaker performs better in both scenarios, indicating that its effectiveness is not due to the multi-tower structure, which utilizes user features in a more fine-grained way. Instead, it is effective because it removes shortcut learning on the user side, validating our motivation. Additionally, the comparison with Breaker$_{2-}$ validates the effectiveness of the delayed parameter update strategy.

\end{itemize}
\subsubsection{Efficiency Comparison} We compare the model sizes and runtime performances of different methods in Table \ref{tab:offline_res}. To ensure consistency with online CPU inference, we conducted our tests on a PC equipped with a 48-core CPU and 232G of memory.
\begin{itemize}
    \item \textbf{Inference Efficiency of Baseline Methods:} Among the baseline methods, gradient-boosting decision tree models like LightGBM boast the fastest inference speed due to their superior model structure and efficient parallel mechanism. For instance, LightGBM outperforms DeepFM in scenario B while being 3 times faster in inference. Methods based on cross and attention structures have more parameters than those based on Factorization Machines (FM). They learn feature interactions through specific network layers, resulting in higher computational complexity and lower inference efficiency. The MultiCR method, like Breaker, employs a multi-tower structure to learn user-item preferences without feature interactions, resulting in lower computational complexity and higher inference efficiency compared to DeepFM.

    \item \textbf{Breaker's Superior Inference Efficiency and Deployability:} In both scenarios, the Breaker model demonstrated highly competitive inference efficiency, achieving approximately a 40\% faster inference speed compared to the best-performing neural network models, DeepFM or MultiCR, in the baseline methods. Additionally, it has fewer parameters, which is beneficial for online deployment.
\end{itemize}

\begin{figure}[t]
  \centering
  \includegraphics[width=\linewidth]{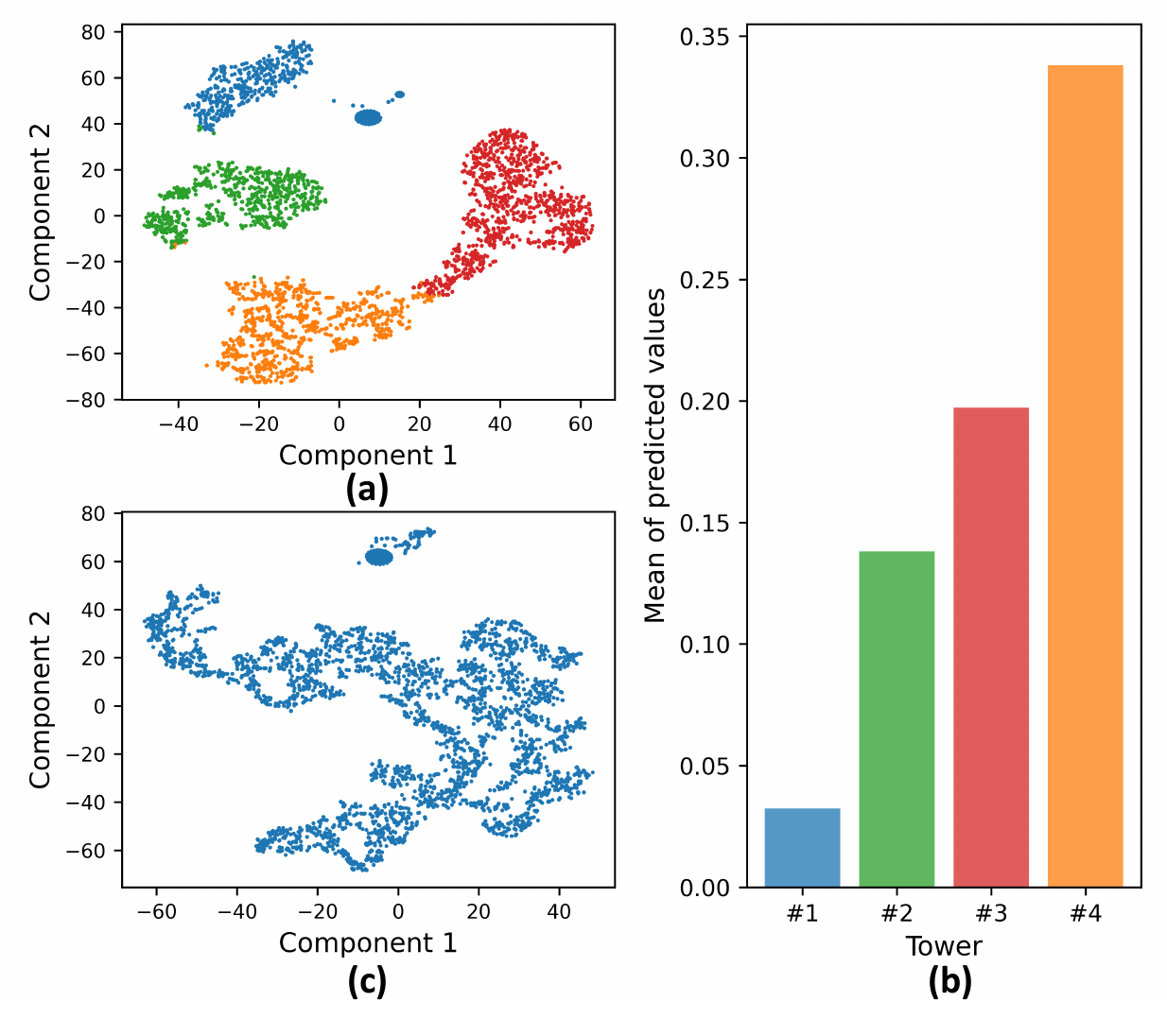}
  \caption{(a) The t-SNE clustering visualizes user representations with various colors indicating cluster labels predicted by URCM. (b) The average predicted values $\overline{\hat{y}_c^k}$ from each tower reflect the different intrinsic tendencies of users within each cluster. (c) The t-SNE clustering showcases user representations extracted by the Breaker$_{-1}$ model, which excludes the auxiliary clustering task.}
  \label{fig:expr_clus}
  \Description{}
\end{figure}

\subsection{In-depth Analysis}
To gain a deeper understanding and analysis of our model, we propose the following four research questions (RQs).
\begin{itemize}
    \item \textbf{RQ1:} How does the Breaker perform in terms of user clustering?
    \item \textbf{RQ2:} How does the number of clusters affect the model performance?
    \item \textbf{RQ3:} How can the Breaker remove the shortcut cues of the user's intrinsic tendencies?
     \item \textbf{RQ4:} Does cluster-driven multi-tower learning improve fine-grained user-item preferences?
\end{itemize}
\subsubsection{Performance of Breaker in User Clustering (RQ1)}\label{4_3_1}
We randomly selected 3,000 samples from the test sets of two scenarios for analysis. We visualized the user representations derived from the URCM module using t-SNE \cite{van2008visualizing}, as shown in Figure \ref{fig:expr_clus}. From the visualization, we can obtain the following inspiring observations in terms of user representation and cluster assignment:
\begin{itemize}
    \item \textbf{Impact of Clustering Task on User Representation:} Figure \ref{fig:expr_clus}(c) shows the results from the ablation model that excludes the auxiliary clustering task. By comparing Figure \ref{fig:expr_clus}(a) and (c), it is evident that the clustering task facilitates the learning of a representation specifically for clustering. To further verify, we also employed the K-means clustering algorithm on the user representations obtained from both models. As indicated in Table \ref{tab:cluster_res},  the silhouette coefficient of Breaker is substantially higher than that of Breaker$_{1-}$ in both scenarios, suggesting that the clustering task has yielded user representations with more distinct clustering properties. Moreover, Figure \ref{fig:expr_clus}(b) illustrates the average predicted values of each tower, which exhibits significant differences. This indicates that the clusters corresponding to each tower have varying levels of user intrinsic tendencies.
    \item \textbf{Clustering Performance:} In Figure \ref{fig:expr_clus}(a), different colors denote the cluster assignment, which aligns closely with the underlying data structure. Across both scenarios, when comparing the cluster assignment results of the K-means algorithm with those from Breaker, the silhouette coefficients are nearly identical as shown in Table \ref{tab:cluster_res}. This consistency suggests that Breaker not only enhances user representation learning but also achieves precise cluster assignment.
\end{itemize}

\begin{table}[t]
\caption{Comparison of silhouette coefficients for user representations from Breaker and Breaker$_{1-}$, clustered by K-Means and Breaker.}
\label{tab:cluster_res}
\begin{tabular}{c|cc|cc}
\toprule
\multirow{1}{*}{Scenario} & \multicolumn{2}{c|}{A} & \multicolumn{2}{c}{B} \\
\multirow{1}{*}{User Rep. from}  & Breaker     & Breaker$_{1-}$     & Breaker     & Breaker$_{1-}$    \\ \midrule
K-Means                   & 0.97        &    0.58           & 0.63        &      0.43        \\
Breaker                  & 0.97           &    -           & 0.62       &       -       \\ \bottomrule
\end{tabular}
\end{table}

\subsubsection{The Influence of Cluster Number on Model Performance (RQ2)}
\begin{figure}[t]
  \centering
  \includegraphics[width=\linewidth]{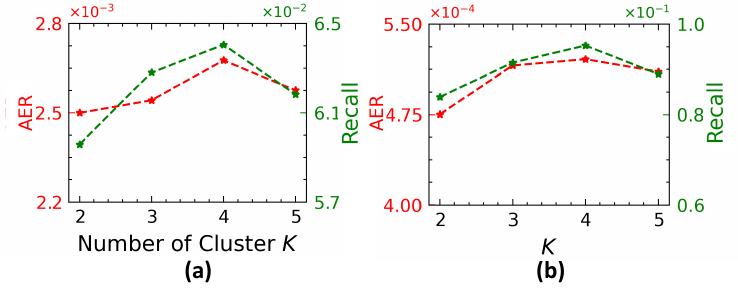}
  \caption{The impact of the number of clusters on scenario A (as shown in part a) and on scenario B (as shown in part b).}
  \label{fig:hyper_param}
  \Description{}
\end{figure}
We have analyzed the influence of cluster numbers on the performance of the Breaker model. As shown in Figure \ref{fig:hyper_param}, an overly large or small number of clusters tends to degrade the model's performance in both scenarios. The model performs optimally when the number of clusters is set to 4.
We also noticed that the number of clusters has a more pronounced impact on the model's performance in scenario B. This can be partially attributed to the sparse interaction data and the scarcity of positive samples in the single-slot recommendation system, leading to inherent volatility in the metrics. Nevertheless, in most cases, it still surpasses the baseline method.

\subsubsection{Eliminating Shortcut Cues of Intrinsic User Tendencies (RQ3)}

\begin{figure}[t]
  \centering
  \includegraphics[width=\linewidth]{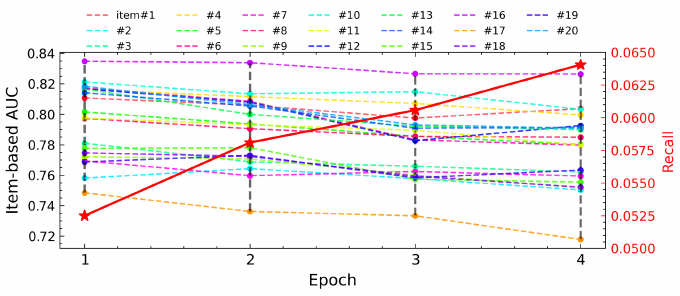}
  \caption{The dashed line represents the item-based AUC, which calculates the AUC for samples within each item. The solid red line represents the recall@1 metric. The drop in item-based AUC suggests increased learning difficulty on the user side, forcing the model to focus on user-item preferences, leading to higher recall metrics.}
  \label{fig:removing}
  \Description{}
\end{figure}

To assess the model's performance in eliminating shortcut cues during training, we used the item-based AUC metric, which calculates the AUC values for samples under each item. The metric removes the influence of different items and indirectly reflects the model's ranking ability for users' intrinsic tendencies.
Figure \ref{fig:removing} shows that the item-based AUC metric declines for most items as training progresses. This is because the model clusters users with similar intrinsic tendencies together, making it difficult to capture the differences in users' intrinsic tendencies.
Conversely, the recall metric improves with training, indicating that the model is getting better at learning user-item preferences. This confirms the effectiveness of Breaker, which increases the task's complexity in learning users' intrinsic tendencies, forcing it to concentrate on user-item preferences, thereby mitigating shortcut bias.

\subsubsection{Learning Fine-grained User-item Preferences (RQ4)}

\begin{figure}[t]
  \centering
  \includegraphics[width=\linewidth]{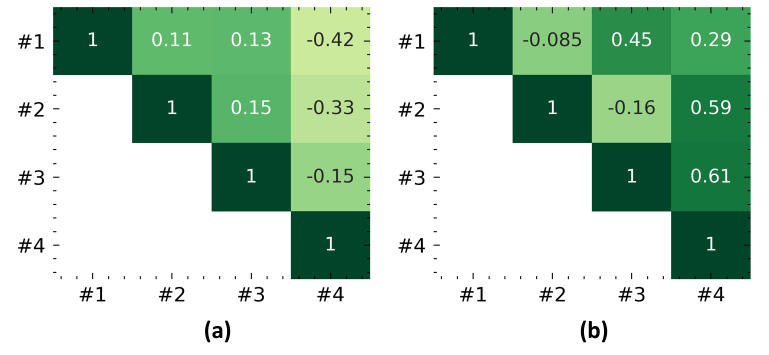}
  \caption{The correlation coefficient between prediction values across towers for scenarios A (a) and B (b) is generally low, suggesting that the multi-tower effectively captures diverse fine-grained user-item preferences.}
  \label{fig:off_expr}
  \Description{}
\end{figure}

In the CPMM, the item representation is concatenated with the user representation and fed into each tower to obtain matching scores. To evaluate the effectiveness of the multi-tower structure, we get the matching score matrix $\hat{Y}_c$, with dimensions $N_t\times K$, where $N_t$ is the sample size. Subsequently, we compute the correlation coefficients between each column of this matrix, as shown in Figure \ref{fig:off_expr}. In particular, we observe that the correlation coefficient matrix exhibits relatively low values, suggesting that each tower can capture the fine-grained user-item preferences.

\subsection{Performance of Online A/B Tests}
As shown in Table \ref{tab:offline_res}, the requirement for efficient model inference in our online system makes it infeasible to deploy all baseline models.

\begin{figure}[t]
  \centering
  \includegraphics[width=\linewidth]{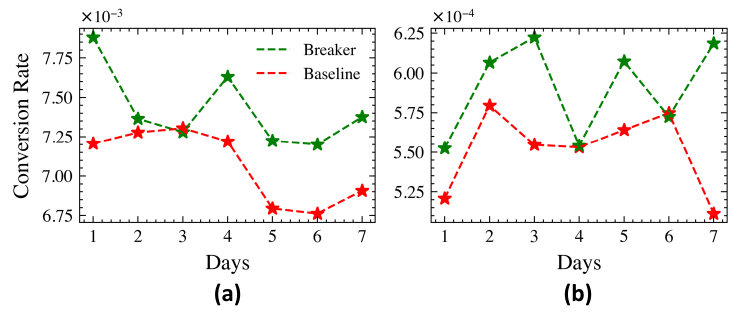}
  \caption{The online A/B test results on scenario A (as shown in part a) and scenario B (b). Considering model efficiency, it is not feasible to deploy all baseline models online. In scenario A, the baseline method employed is DeepFM, whereas, in scenario B, the corresponding baseline is LightGBM.}
  \label{fig:abtest}
  \Description{}
\end{figure}

Instead, we have successfully implemented LightGBM, DeepFM, and Breaker within our online system, supporting the targeted marketing display recommendation in Meituan Payment service, as depicted in Figure \ref{fig:framework}. To ensure a fair comparison, we have controlled the A/B testing to involve homogeneous requests, with each model serving millions of users daily. The A/B test was conducted for 7 days in March 2023. As shown in Figure \ref{fig:abtest}, Breaker significantly improved the conversion rate by 4.93\% compared to DeepFM in scenario A (with a
$p$-value of 0.0039 from the paired samples t-test), and by 7.15\% compared to LightGBM in scenario B (with a $p$-value of 0.0174).

\section{Related Work}
\textbf{Recommendation System}: Recommendation systems predict user-item preferences and suggest relevant items \cite{zhang2019survey}. They typically include embedding, feature interaction, representation, and prediction layers. The embedding layer converts discrete features into continuous vectors for better expression, as seen in XcrossNet \cite{yu2021xcrossnet} and AutoDis \cite{guo2021embedding}. The feature interaction layer improves the expressiveness of the model by capturing feature interactions, as demonstrated by DeepFM \cite{guo2017deepfm}, AutoINT \cite{song2019autoint}, and EDCN \cite{chen2021enhancing}. The feature representation layer learns vector-level weights to address fixed feature representation issues, as seen in FiBiNET \cite{huang2019fibinet} and FRNet \cite{wang2022enhancing}. Debiasing research tackles biases such as selection, consistency, exposure, position, and popularity \cite{chen2023bias}. However, shortcut bias remains unexplored in single-slot recommendation systems. While \cite{lyu2022see} may be relevant, it does not introduce inductive biases to remove shortcut biases. Consequently, even if user intent and item preference are decoupled, it fails to debias effectively.

\textbf{Shortcut Bias}: DNNs often exhibit shortcut bias by learning the simplest solution to a problem, which may not align with the desired generalization \cite{geirhos2020shortcut}. This occurs because DNNs focus on minimizing training loss rather than capturing the task's underlying complexity. To address this, methods commonly increase task difficulty, encouraging the model to thoroughly explore the solution space and improve generalization \cite{wad2022equivariance, fei2022xcon, sohoni2020no}. Breaker also adopts this strategy to mitigate user-side shortcut biases by optimizing within clusters of similar users. This increased learning difficulty on the user side, forcing the model to focus on user-item preferences.

\section{Conclusion}
In this paper, we propose Breaker for single-slot recommendation systems. By incorporating an auxiliary clustering task for user representation, Breaker eliminates the shortcut bias caused by the user's intrinsic tendencies, enabling it to learn user-item preferences more accurately and comprehensively. Our offline and online experiments demonstrate that Breaker outperforms existing baseline methods. Moreover, Breaker has been successfully deployed, serving tens of millions of users daily.

\bibliographystyle{ACM-Reference-Format}
\balance

\end{document}